\documentclass[aps,prl,showpacs,twocolumn,superscriptaddress, floatfix, tightenlines, amsmath, amssymb,longbibliography]{revtex4-1}
\usepackage[pdftex]{graphicx} 

\usepackage{natbib}
\usepackage{booktabs}
\usepackage{color}
\usepackage{url}
\usepackage{float}
\usepackage{amsmath}
\usepackage{soul,xcolor}
\usepackage{hyperref}
\usepackage{nameref}
\newcommand{\HLIO}{H$_3$LiIr$_2$O$_6$} 
\newcommand{\Li}{$\alpha$-Li$_2$IrO$_3$}
\newcommand{\Na}{Na$_2$IrO$_3$}
\DeclareUnicodeCharacter{2009}{\,} 
\begin{document}

\title{Momentum-independent magnetic excitation continuum in the honeycomb iridate H$_3$LiIr$_2$O$_6$}

\author{A. de la Torre}
\email[Corresponding author:]{adlt@brown.edu}
\affiliation{Department of Physics, Brown University, Providence, Rhode Island 02912, United States}
\author{B. Zager}
\affiliation{Department of Physics, Brown University, Providence, Rhode Island 02912, United States}
\author{F. Bahrami}
\affiliation{Department of Physics, Boston College, Chestnut Hill, MA 02467, USA}
\author{M. H. Upton}
\affiliation{Advanced Photon Source, Argonne National Laboratory, Argonne, Illinois 60439, USA}
\author{J. Kim}
\affiliation{Advanced Photon Source, Argonne National Laboratory, Argonne, Illinois 60439, USA}
\author{G. Fabbris}
\affiliation{Advanced Photon Source, Argonne National Laboratory, Argonne, Illinois 60439, USA}
\author{G.-H. Lee}
\affiliation{Advanced Light Source, Lawrence Berkeley National Laboratory, Berkeley 94720, USA}
\author{W.Yang}
\affiliation{Advanced Light Source, Lawrence Berkeley National Laboratory, Berkeley 94720, USA}
\author{D. Haskel}
\affiliation{Advanced Photon Source, Argonne National Laboratory, Argonne, Illinois 60439, USA}
\author{F. Tafti}
\affiliation{Department of Physics, Boston College, Chestnut Hill, MA 02467, USA}
\author{K. W. Plumb}
\email[Corresponding author:]{kemp_plumb@brown.edu}	
\affiliation{Department of Physics, Brown University, Providence, Rhode Island 02912, United States}

\date{\today}

\maketitle

\textbf{In the search for realizations of Quantum Spin Liquids (QSL), it is essential to understand the interplay between inherent disorder and the correlated fluctuating spin ground state. \HLIO{} is regarded as a spin liquid proximate to the Kitaev-limit (KQSL) in which H zero-point motion and stacking faults are known to be present. Bond disorder has been invoked to account for the existence of unexpected low-energy spin excitations. Controversy remains about the nature of the underlying correlated state and if any KQSL physics survives. Here, we use resonant X-ray spectroscopies to map the collective excitations in \HLIO{} and characterize its magnetic state. We uncover a broad bandwidth and momentum-independent continuum of magnetic excitations at low temperatures that are distinct from the paramagnetic state. The center energy and high-energy tail of the continuum are consistent with expectations for dominant ferromagnetic Kitaev interactions between dynamically fluctuating spins. The absence of a momentum dependence to these excitations indicates a broken translational invariance. Our data support an interpretation of \HLIO{} as a disordered topological spin liquid in close proximity to bond-disordered versions of the KQSL. Our results shed light on how random disorder affects topological magnetic states and have implications for future experimental and theoretical works toward realizing the Kitaev model in condensed matter systems.}


Quantum Spin Liquids (QSLs) encompass a rich family of phases of matter with dynamically fluctuating and long-range entangled spins at zero temperature \cite{savary_quantum_2017,broholm_quantum_2020}. One particularly important QSL realization is the Kitaev model (KQSL). This exactly soluble model consists of highly-frustrated bond directional Ising-like interactions between spin-1/2 on the two-dimensional Honeycomb lattice and can host topologically protected fractionalized excitations \cite{KITAEV20062}. Material realizations of this model have been proposed in strong spin-orbit coupled $4d/5d$ transition metal ($TM$) compounds with edge sharing $TML_6$ ($L = $ O, Cl) octahedra in a Honeycomb lattice \cite{PhysRevLett.102.017205}. In this case, the Heisenberg, $J$, and symmetric off-diagonal, $\Gamma$, exchange terms between $J_{eff} = 1/2$ pseudospins are predicted to be suppressed in favor of finite anisotropic exchange, $K$ \cite{Rau2014}. However, the most heavily studied candidate Kitaev materials, \Li{}, \Na{} \cite{B207282C,Gegenwart_Na,Gegenwart_TN_Li_Na} and $\alpha$-RuCl$_3$\cite{Plumb2014,Banerjee2016}, all magnetically order due to the existence of additional exchange interactions of comparable magnitude to $|K|$, as a result of non-cubic distortions of the $TML_6$ octahedra and $4d/5d$ orbital extent\cite{Rau_review,Takagi2019}. Nonetheless, measurements of the excitation spectrum in these compounds have revealed ca high-energy magnetic continuum consistent with dominant Kitaev nearest-neighbor bond-directional interactions \cite{Banerjee2016,chun_optical_2021,kim_dynamic_2020,hwan_chun_direct_2015,revelli_fingerprints_2020}. 

\begin{figure*}[!ht]
    \includegraphics[width=\textwidth]{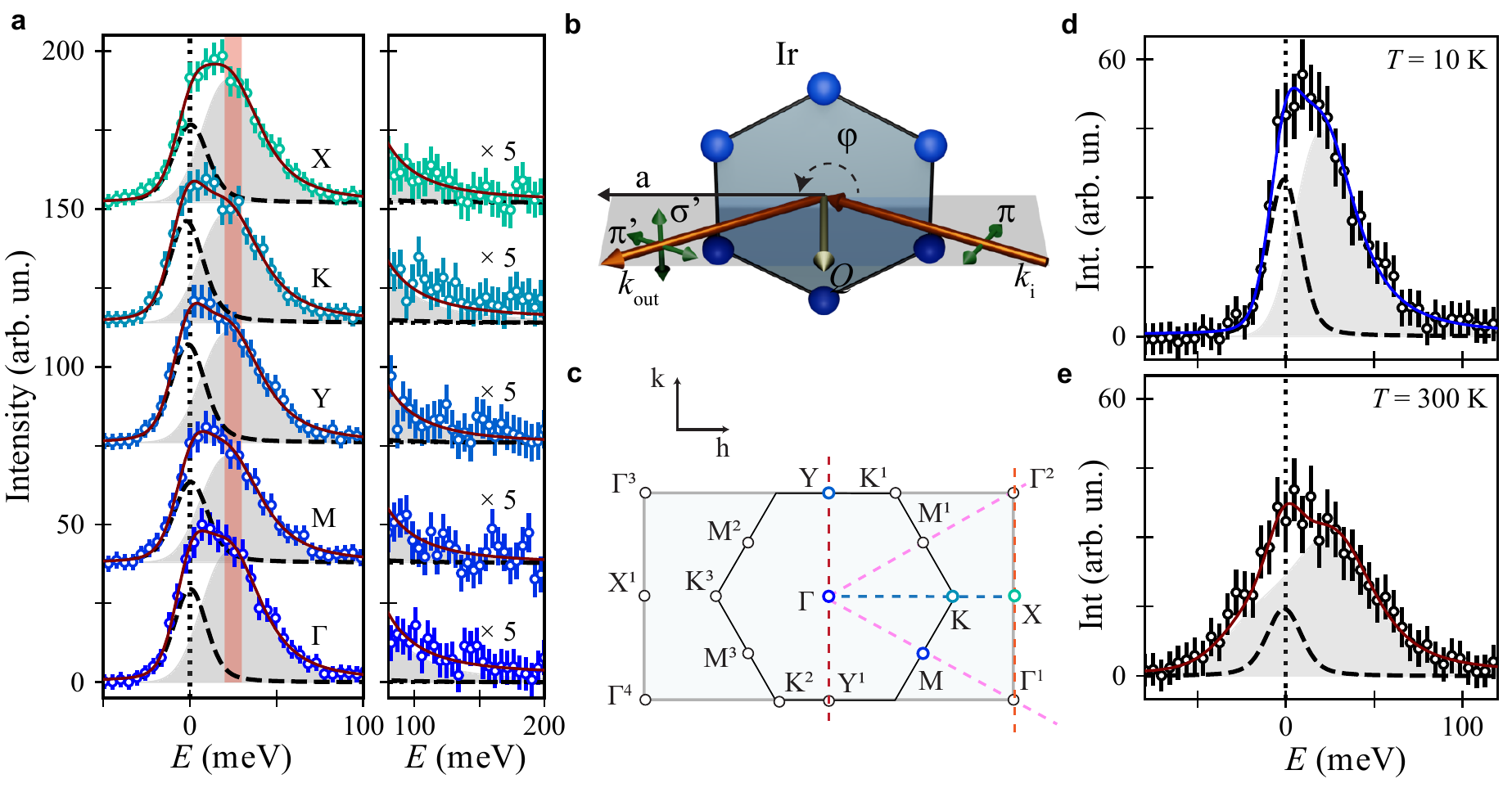}
	\caption{\textbf{High-resolution resonant inelastic X-ray spectra of \HLIO{} at the Ir $L_3$ edge.} \textbf{a}, RIXS spectra at $T = 10$~K at high symmetry points of the Brillouin zone (BZ) (circular markers). Black solid line is a fit including a Voigt profile for the elastic line (dashed black line) and a damped harmonic oscillator (grey shading) centered at $E_0 = 25 \pm 5$~meV (red bar of width $10$~meV). Spectra have been shifted vertically for clarity. \textbf{b}, Sketch of the scattering geometry. The hexagonal arrangement of blue spheres represent a honeycomb layer of Ir$^{4+}$ ions within the monoclinic crystal structure of \HLIO{}. The incident (\textbf{$k_i$}) and outgoing (\textbf{$k_f$}) radiation (orange arrows) define the scattering plane (grey). Green arrows show the polarization of the X-ray electric field ( $\pi$: in-plane; $\sigma, \sigma^{\prime{}}$: out-of- plane for the incoming and outgoing X-ray beam respectively). $\varphi$ is the azimuthal angle defined by the crystallographic a-axis and the scattering plane. The data in panel \textbf{a} was taken at $\varphi = 180 ^{\circ}$. \textbf{c}, Schematic of the extended hexagonal BZ highlighting relevant symmetry point and directions. $L$ varies between 5.91 and 5.95 r.l.u. \textbf{d}, RIXS spectra at $\Gamma$ measured at $T = 10$~K and, \textbf{e}, $T = 300$~K and $\varphi = 0^{\circ}$. Red and blue solid lines are fit to the data as described above.}
	\label{fig:f1}
\end{figure*}

\HLIO{} is the most salient example of a new generation of Kitaev compounds. It is synthesized from the parent compound \Li{} through the replacement of inter- honeycomb layer Li with H \cite{molecules27030871}. This leads to a reduction of inter-LiIr$_2$O$_6$ layer coordination from octahedral to linear and a consequent modification of the intra-layer Ir-O-Ir bond angles and Ir-Ir bond distance. As a result, superexchange pathways are modified with respect to those of \Li{} and a different magnetic state is expected in \HLIO{} \cite{molecules27030871,bette_solution_2017}. Temperature dependent measurements of the magnetic susceptibility in \HLIO{} show no evidence for long range magnetic order down to 5~mK, despite a Curie-Weiss temperature $\theta_{CW} \approx - 105$ K, as confirmed by the NMR Knight shift \cite{kitagawa_spinorbital-entangled_2018,Lee_NMR_2023}. The NMR relaxation rate rules out a spin glass in favor of dynamically fluctuating spins \cite{kitagawa_spinorbital-entangled_2018}. Raman spectroscopy \cite{Pei_Raman_2020} finds a continuum of magnetic excitations similar to that in the proximate Kitaev magnet $\alpha$-RuCl$_3$ \cite{Sandilands_raman_rucl3_2015,nasu_fermionic_2016}. However, the observation of a non-zero NMR relaxation rate and $T^{-1/2}$ $C/T$ divergence indicate $E=0$ spin excitations that are not consistent with expectations for a  pure Kitaev QSL state \cite{kitagawa_spinorbital-entangled_2018}. Thus, the nature of the unconventional magnetic ground state and the associated collective excitations in \HLIO{} remains unknown.

\begin{figure*}[!ht]
    \includegraphics[width=\textwidth]{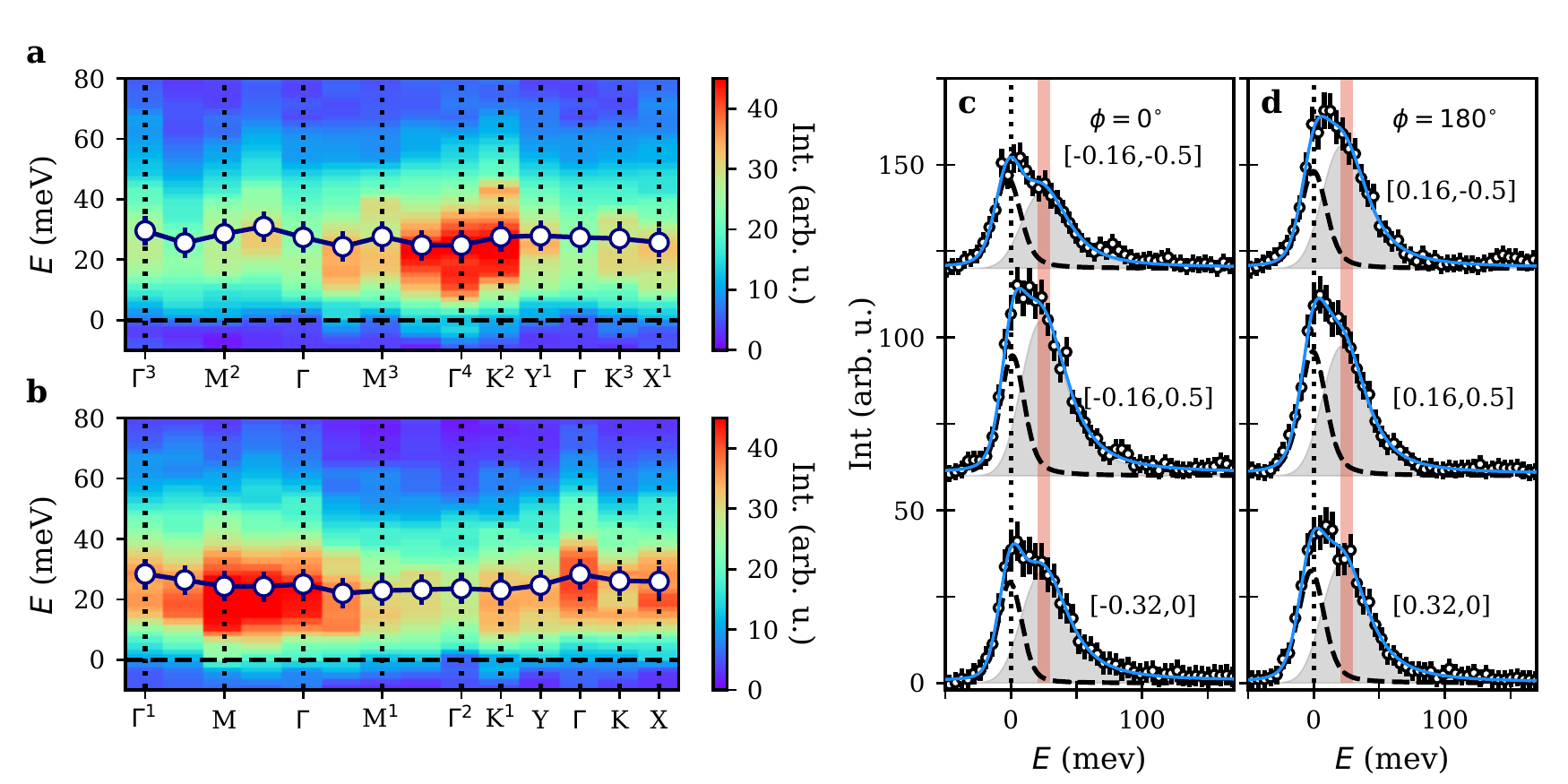}
	\caption{\textbf{Absence of short range correlations in \HLIO{}.} \textbf{a} High-resolution elastic-background-subtracted RIXS intensity along high symmetry paths with $H \leq 0$, $\varphi = 0^{\circ}$ and, \textbf{b}, $H \geq 0$, $\varphi = 180^{\circ}$ . The overlay circular markers indicate the center position of the continuum of magnetic excitations as extracted from a fit to the data. \textbf{c}, \HLIO{} $T = 10$~K RIXS intensity at the wavevectors of the $120^{\circ}$ spiral order of \Li{} for $\varphi = 0^{\circ}$ and, \textbf{d}, $\varphi = 180^{\circ}$ . Blue solid line is a fit to the data including a Voigt profile for the elastic line (dotted black line) and a damped harmonic oscillator (grey shading) centered at $E_0 = 25$~meV (red bar of width $10$~meV).}
	\label{fig:f2}
\end{figure*}

Achieving a microscopic description of the magnetic state in \HLIO{} is complicated by disorder. Quasi-static H zero point motion \cite{geirhos_quantum_2020, wang_possible_2020,Lee_NMR_2023}, present even in crystallographic pristine samples, can lead to local variations of the intra-layer exchange interactions \cite{li_role_2018, yadav_strong_2018}. Furthermore, the heavily faulted stacking structure due to the linear inter-layer O-H-O coordination might also result on random intra-layer magnetic exchanges \cite{bette_solution_2017, molecules27030871}. It remains to be understood how the random distribution of magnetic exchanges modify the magnetic Hamiltonian to render a state with no long-range magnetic order or frozen moments but with low-temperature low-energy spin excitations. One possibility, is that bond-disorder brings \HLIO{} away from the QSL regime \cite{li_role_2018, yadav_strong_2018} into a state where the formation of a long-range magnetically order phase is inhibited but with short-range correlations remnant of ordered states in \Li{} and \Na{} \cite{hwan_chun_direct_2015,chun_optical_2021}. A second possibility to account for the absence of frozen moments, is that bond-disorder suppresses Kitaev exchange in favor of a random distribution of nearest neighbor $J$ promoting the formation of a random valence bond quantum paramagnet \cite{yamaguchi_randomness-induced_2017}. In these two scenarios, the low-energy excitations observed in \HLIO{} could be explained by the presence of fluctuating unpaired spins that give rise to the $C/T$ scaling \cite{Spin_singlet_Lee, kimchi_scaling_2018}. Within a third alternative, the thermodynamic observations in \HLIO{} can be explained by bond-disordered extensions of the pure ($J=0, \Gamma =0$) and extended Kitaev QSL (BD-KQSL). In this case, bond-disorder acts to pin a random distribution of flux degrees of freedom in the underlying KQSL and leads to the divergent low-energy density of states and associated $T^{-1/2}$ low-temperature specific heat \cite{slagle_theory_2018,knolle_bond-disordered_2019, kitagawa_spinorbital-entangled_2018,Nasu_thermodynamic,KAO2021168506,Vacancy_Induced_PRX_Perkins,Andrade_PRL_2022}. We remark that the disordered Kitaev state is proximal to the pristine KQSL, but it represents a new disordered phase of topological matter. All of these models represent distinct states that account for the absence of frozen moments and can explain the thermodynamic observations in \HLIO{}. However, they can be distinguished by their high-energy collective excitations as measured by dynamical spin-spin correlations, $S (q,\omega)$. For example, the proximity to magnetically ordered phases might lead to correlations on longer length scales than the nearest-neighbor only correlations of the BD-KQSL  \cite{hwan_chun_direct_2015,chun_optical_2021,knolle_dynamics_2014,knolle_dynamics_fractionalization_2015}. Similarly, singlet-triplet excitations at the scale of $J$ in a random valence bond state lead to a characteristic momentum dependence of $S (q,\omega)$, with a maximum of intensity at the zone boundary and lack of intensity at the zone center \cite{zhu_disorder-induced_2017,Kimchi_YbMgGaO4},  distinct from the BD-KQSL \cite{knolle_dynamics_beyond_2018,knolle_bond-disordered_2019}. A momentum resolved spectroscopic measurement of the spin excitation spectrum is essential to distinguish between these two opposing views, discern the role of disorder, and reveal the magnetic ground state of \HLIO{}.

\begin{figure*}[t]
	\includegraphics{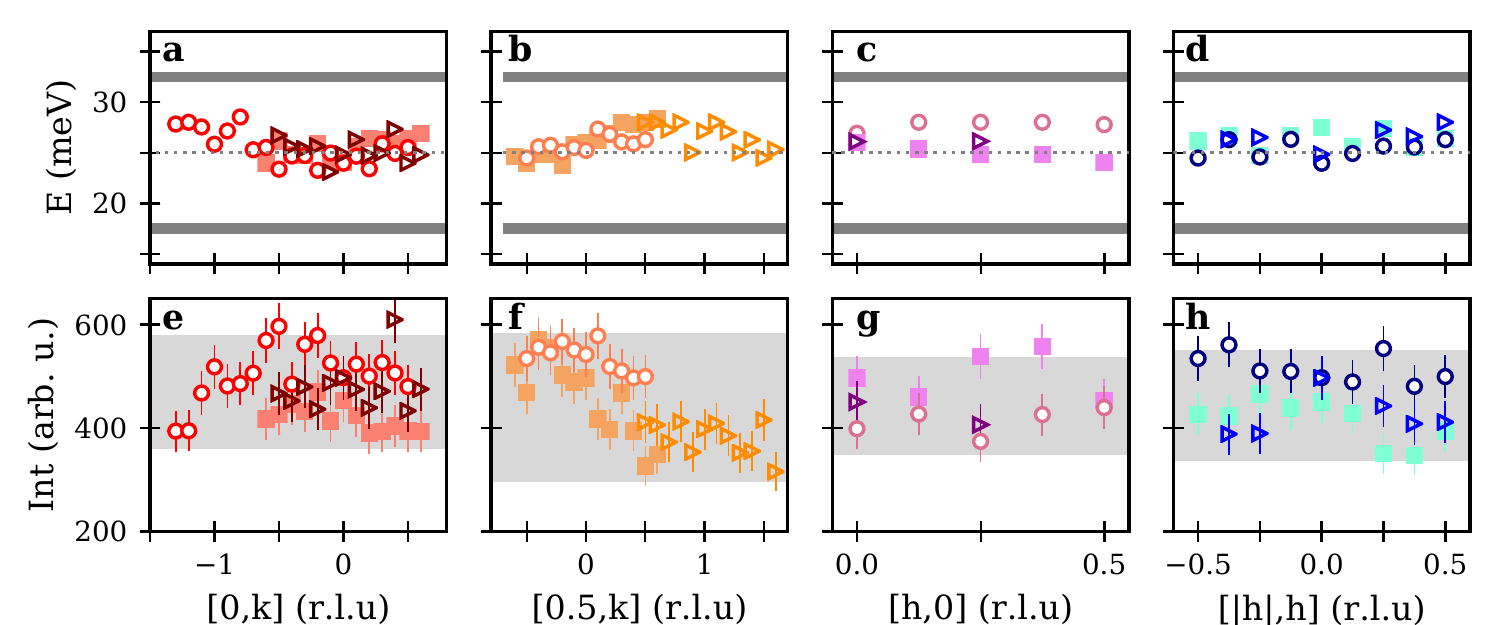}
	\caption{\textbf{Momentum-independent magnetic continuum.} \textbf{a-d} Center energy of the continuum of magnetic excitations across the Brillouin Zone as extracted from a fit to the inelastic RIXS intensity to a damped harmonic oscillator. The dashed line at $E = 25$~meV is the average momentum position of the continuum. The energy resolution was relaxed to a $FWHM = 28$~meV as indicated by the solid grey lines.\textbf{e-h} Integrated elastic-background-subtracted RIXS intensity in the range $E \in [-30,170]$~meV. The grey shaded rectangle represents the variation of the elastic intensity. Circular markers ($\varphi = 0^{\circ}$), triangular markers ($\varphi = 180^{\circ}$) and square markers ($\varphi = 90^{\circ}$) show the azimuthal dependence.} 
	\label{fig:f3}
\end{figure*}

The small volume of available \HLIO{} single crystals (see the sample growth and characterization section and RIXS methods sections), high neutron absorption cross-section of Ir, and the incoherent scattering from H inhibits inelastic neutron scattering from accessing the magnetic spectrum in this compound \cite{Mourigal2013,Tian-Heng_herbertsmithite_2016, Banerjee2016,Plumb2019}. High resolution resonant inelastic X-ray scattering (RIXS) at the Ir $L_3$ edge offers an alternative. In Fig.\ref{fig:f1} \textbf{a} we show the RIXS intensity of single crystals of \HLIO{} at the Ir L$_3$ ($E_i = 11.215$~keV) with state-of-the-art energy resolution (full width half maximum FWHM $=$ 20~meV) at high symmetry points of the pseudo-honeycomb lattice at $T = 10$~K following the experimental geometry sketched in Fig.\ref{fig:f1} \textbf{b}. Our measurement temperature is one order of magnitude smaller than $\theta_{CW}$ and below the ordering temperature of pristine \Li{} \cite{Gegenwart_TN_Li_Na}. The RIXS scans reveal a broad, $\sigma =40 \pm 5$~meV bandwidth, inelastic signal centered around $E = 25 \pm 5$~meV. This signal resonates at the Ir L$_3$ edge (see Supplementary information) and is well described by an overdamped harmonic oscillator (DHO) weighted by a Bose factor ($1/(1-e^{-E/k_BT})$) and convolved with a Gaussian resolution \cite{Paramagnon_rixs}. The tail of the DHO extends into a high energy magnetic continuum spanning up to $E = 170$~meV. Given the absence of any sharp features over a bandwidth that is 8 times larger than the energy resolution of our measurement, we attribute this signal to a continuum of magnetic excitations. The inelastic spectrum is remarkably similar at all high symmetry points, being of equal or comparable intensity to the elastic line and without an evident dispersion or intensity modulation. The temperature dependent RIXS intensity Fig.\ref{fig:f1} \textbf{d,e} does not show any sharp variations that would indicate a magnetic transition. We find that the inelastic intensity smoothly varies and follows directly from detailed-balance of the harmonic oscillator component. However, at $T \!=\! 300\,\, \mathrm{K}\! \approx\! 3\theta_{CW}$, the DHO is centered at $E = 38 \pm 5$~meV with $\sigma = 45 \pm 5$~meV bandwidth. This hardening is suggestive of a spectral weight transfer between multiple modes that are unresolvable with our energy resolution. Thus, our data shows a collective excitation spectrum in \HLIO{} that is dominated by the presence of strong overdamped magnetic fluctuations at low temperature and that is different from the paramagnetic state above $\theta_{CW}$. The absence of a momentum dependence is contrary to expectations for a random distribution of singlets in a valence bond solid with dominant nearest neighbor correlations \cite{zhu_disorder-induced_2017,Kimchi_YbMgGaO4} and disordered Heisenberg AFM models \cite{PhysRevB.92.134407} that predict a vanishing intensity at $\Gamma$ points.  

One possible explanation for the excitations we observe is that bond-disorder and stacking faults act to inhibit an underlying magnetically ordered phase in \HLIO{}\cite{li_role_2018,yadav_strong_2018}. In this scenario, remnant dynamical correlations reflecting a disorder limited short range order should be present and give rise to a characteristic momentum dependence of $S(q,\omega)$. Good points of comparison are momentum-resolved measurements of magnetic X-ray scattering in \Li{} and \Na{} above their magnetic transition temperatures. For both compounds, the diffuse magnetic scattering is characterized by an intensity modulation that peaks at their respective ordering vectors and depends on the projection of the incident X-ray polarization on spins thermally-fluctuating about an average orientation \cite{hwan_chun_direct_2015,chun_optical_2021}. Similar information can be extracted by examining the magnetic RIXS response as a function of momentum and azimuthal angle $\varphi$ as sketched in Fig. \ref{fig:f1} \textbf{c} and \textbf{d}. We extract the inelastic RIXS intensity by subtracting the elastic line contribution determined from a fit to a resolution limited Voigt profile.  Fig. \ref{fig:f2} \textbf{a}  shows the momentum dependence of inelastic magnetic intensity along high symmetry directions covering multiple zone center ($\Gamma$) and zone boundary ($M$, $K$, $Y$ and $X$) points with  $H\leq 0$ and $\varphi = 0$. Equivalent points with $H \geq 0$ and $\varphi = 180^{\circ}$ are shown in Fig. \ref{fig:f2} \textbf{b}. For both azimuthal configurations, the excitation spectra has comparable intensity at the zone center and zone boundary, it is non-dispersive, and devoid of sharp coherent modes. This data rules out any short range zig-zag correlations as present in \Na{} that lead to strong modulation of the magnetic diffuse scattering at $M$, $Y$ and $X$ \cite{hwan_chun_direct_2015,kim_dynamic_2020}. Moreover, in Fig. \ref{fig:f2} \textbf{c,d} we show the RIXS spectra at $q_{\alpha} = [\pm 0.16, -0.5], [\pm 0.16, 0.5],[\pm0.32, 0]$, the ordering vectors for the three $120^{\circ}$ domains of the incommensurate spiral order in \Li{} \cite{chun_optical_2021}.  The inelastic data across all momentum points and azimuthal angles can be fit to the same functional form as for the high symmetry points in Fig.\ref{fig:f1} without any additional broadening or energy shift to the DHO and without the need to account for the gapless acoustic magnon at $q_{\alpha}$ observed in \Li{} below $T_N = 15$~K \cite{chun_optical_2021}. We thus rule out a disorder limited correlations remnant of an ordered state and conclude that the spin excitations in \HLIO{} are intrinsically different to that of the parent compound or \Na{}. 

\begin{figure*}[!ht]
 \begin{center}
	\includegraphics[width=\textwidth]{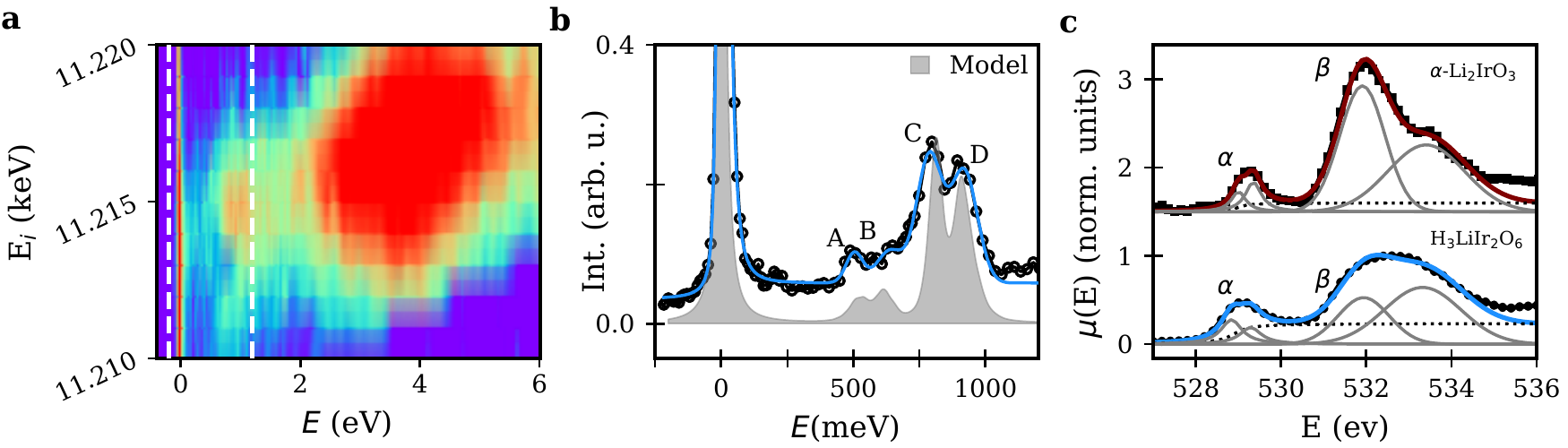}
	\caption{\textbf{Local Ir electronic structure}. \textbf{a}, RIXS intensity as a function of the incident X-ray energy near the Ir $L_3$ edge. The intense inelastic feature centered at $E \approx 10Dq \approx 3.8$~eV is from transitions between occupied $t_{2g}$ states and empty $e_g$ orbitals in a Ir$^{4+}$ ion (see Supplementary information).\textbf{b}, Intra-$t_2g$ RIXS excitations at $E_i = 11.215$~keV (circular markers) compared to the calculated RIXS intensity from exact diagonalization calculations. Blue solid line is a fit to the data including a Voigt peak, aDHO, four Gaussian peaks and an arctan step to account for the background (A-D). \textbf{c}, O K-edge X-ray absorption spectroscopy data and fit to the data in \HLIO{}(circular markers; blue line) and \Li{} (square markers; orange line). Solid grey lines show the two Lorentzian and two Gaussian profiles included in the fit. Dashed line represents the arc tangent step to account for the electron hole continuum.  White dashed lines in \textbf{a} correspond to the energy range shown in \textbf{b}. }
	\label{fig:f4}
	\end{center}
\end{figure*}

In Fig. \ref{fig:f3} \textbf{a-d} we plot the dependence of the center energy of the DHO along four additional high symmetry directions covering the totality of the BZ for three different values of $\varphi$. The inelastic intensity integrated  after subtraction of the elastic line in the energy range $E\!\in\![-30,170]$~meV is shown in Fig. \ref{fig:f3} \textbf{e-h}. We observe no dispersion of the broad continuum centered at $E = 25 \pm 10$~meV (Experimental resolution $FHWM = 28$~meV). The slight intensity modulation in our momentum dependent data does not follow any of the structural symmetries of \HLIO{} and it is correlated variations in the elastic line intensity that are difficult to decouple given our experimental resolution (See Supplementary Material). As such we assign it to an experimental artifact possibly related to self-absorption effects \cite{revelli_fingerprints_2020}. Our RIXS measurements covering many Brillouin zones, azimuthal angles, and the full energy bandwidth of magnetic excitations in \HLIO{} reveals dynamical correlations that are isotropic and with a featureless momentum dependence of the integrated intensity. 

Our data demonstrates the existence of magnetic excitations at high energies ($25$~meV) in addition to the low energy divergence observed in $C/T$ measurements in \HLIO{}. The appearance of two energy scales is suggestive of calculations of $S(q,\omega)$ for pure, extended and bond-disorder gapless KQSL. Within these models the energy dependence of $S (q,\omega)$ is characterized by an intense low-energy ($E \approx 0$) peak at the energy of the fractionalized fermionic excitations that decays exponentially into a high energy continuum (up to $6J$) \cite{knolle_dynamics_2014,knolle_dynamics_fractionalization_2015,knolle_dynamics_beyond_2018,knolle_bond-disordered_2019}. The momentum dependence shows broad isotropic excitations with an intensity modulation that depends on the sign of $K$. For an antiferromagnetic $K$ the $q$ dependence of the broad magnetic continuum at $E = |K|$ exhibits a shift of spectral weight from $\Gamma$ to the zone boundary, while for the ferromagnetic (FM) Kitaev case a maximum of intensity is expected at $\Gamma$ \cite{knolle_dynamics_2014,knolle_dynamics_fractionalization_2015,knolle_dynamics_beyond_2018}. Calculations of the RIXS response for the Honeycomb Kitaev model distinguish between spin conserving (SC) and spin non-conserving (SNC) excitations channels which correspond to dispersive fractionalized fermionic excitations and bound gauge flux excitations respectively. Within our experimental configuration ($\pi$ incident polarization; $2\theta = 90^{\circ}$ scattering angle) the intensity is dominated by polarization-switching SNC channels leading to a nearly featureless excitation spectrum that resembles that accessible by inelastic neutron scattering measurements \cite{Halsz_RIXS_calc}. However, the lack of a clear peak of intensity at $\Gamma$ and of any momentum dependence across the BZ is suggestive of broken translation invariance due to bond-disorder. Thus, we assign the origin of the broad collective excitations centered at $E = 25 \pm 5$~meV and high-energy continuum extending up to $E = 170$~meV to the existence of dominant bond dependent nearest-neighbor spin-spin correlations with a FM Kitaev exchange, $|K| \approx 25$~meV in a bond-disorderd background. While a random distribution of exchange values due to disorder might lead to an apparently larger value of $K$, the extracted value for $K$ is comparable to that of \Li{} and \Na{} \cite{hwan_chun_direct_2015,revelli_fingerprints_2020,kim_dynamic_2020,chun_optical_2021,Magnaterra_na2iro3} and to that extracted from first principles calculations \cite{li_role_2018,yadav_strong_2018}. We remark that the energy resolution of our RIXS experiment $E\!=\!20$~meV integrates over low-energy modes and prevents us from resolving the $E \approx 0$ divergence predicted in the density of states for BD-KQSL models \cite{knolle_bond-disordered_2019}. Similarly, we cannot exclude additional finite but comparable to $k_B T = 0.87$~meV $J$, $\Gamma$ or any other exchange terms in the magnetic Hamiltonian of \HLIO{} \cite{li_role_2018}. Our RIXS data points to \HLIO{} hosting a unique disordered topological state in close proximity to the KQSL resulting from the interplay of $K$ and bond-disorder \cite{knolle_bond-disordered_2019}. 

We finally comment on the specific effects of H on the local Ir electronic structure. While determining the position of the H atoms is not directly accessible in an X-ray experiment, measurements of the crystal field energies and width provide information about any local disorder on IrO$_6$ octahedra \cite{delatorre_PRM_2022}. In Fig. \ref{fig:f4}\textbf{a} we show the RIXS response of \HLIO{} near the Ir L$_3$ as a function of incident energy, $E_i$, and energy transfer, $E$. We focus our discussion on the $ 0.25 < E < 1.2$ eV range corresponding to the intra-$t_{2g}$ excitations  \cite{Gretarsson_PRL_2015}. In this range, the RIXS spectrum is characterized by a set of four Gaussian peaks (Fig. \ref{fig:f4} (b)) (see Supplementary information) of FHMM , $FWHM_{A-D} = 75-120$~meV, larger but comparable to that observed for \Li{} \cite{Gretarsson_PRL_2015}. We model the crystal field excitations in \HLIO{} (Fig. \ref{fig:f4} (b)), by calculating the RIXS intensity from the exact diagonalization of a model Hamiltonian for an Ir$^{4+}$ including nearest-neighbor hopping and local disorder \cite{delatorre_PRB_2021,delatorre_PRM_2022}. Disorder is encoded by sampling the hopping parameters and the magnitude of the trigonal crystal field, $\delta$, from a normal distribution (see Methods). The data is well described including an O mediating hopping with mean value $t_O = 440 $~meV. This value is well within the range of values extracted from density functional theory for an ideal $C2/m$ structure \cite{li_role_2018}, and is $10\%$ larger than that needed to account for the RIXS spectra of \Li{} (see Methods). The larger value of $t_O$ is suggestive of an increased Ir $t_{2g}$-O $2p$ hybridization as a result of the introduction of $H$. This is substantiated by a qualitative comparison of the O-K edge XAS data between \HLIO{} and \Li{} (see Methods). The intensity and width of the $\alpha$-peak at $E = 529$~eV reflects the degree of hybridization between O ${2p}$ and $t_{2g}$ states \cite{sohn_mixing_2013}, which is $1.16$ times more intense and $1.4$ times broader in \HLIO{} than in \Li{}. Moreover, we find a comparable value of the mean $\delta$ between both compounds ( $\delta \!=\! -48$~meV in \HLIO{} and $\delta \!=\! -50$~meV in \Li{}), in agreement with the average $<2\%$ change in the Ir-O bond angles and $<4\%$ in Ir-Ir bond distance with respect to that of \Li{} \cite{bette_solution_2017}. However, we find that a large distribution of $\delta$ values, as encoded by the standard deviation $\sigma_{\delta} = 20$~meV, is needed to account for the broader crystal field excitations in \HLIO{} and the relative intensity ratio between peaks $A-B$ and $C-D$. This is consistent with the existence of slow H-ion motion at low temperature, which can generate variations of the local IrO$_6$ environment \cite{geirhos_quantum_2020,Lee_NMR_2023}. Thus, our Ir crystal field RIXS data, O XAS measurements and analysis is consistent with the existence of local bond disorder on IrO$_6$ octahedra and points to an enhanced Ir-O hybridization as the leading mechanism favoring a dominant Kitaev-like exchange in \HLIO{}. 

In summary, the spin excitation spectrum of \HLIO{} is characterized by broad isotropic excitations of comparable intensity to the elastic line without a momentum dependence of the integrated intensity. Altogether, our data support an interpretation of the magnetic excitation spectrum of \HLIO{} as emerging from the interplay of disorder and dominant Kitaev-like nearest-neighbor bond-directional interactions between dynamically fluctuating spins. Of the theoretical proposals that have been put forward to explain the thermodynamic properties of \HLIO{}, bond-disorder KQSL models are the closest to describing our observations. However, the lack of a clear intensity modulation across the BZ indicate that these excitations are of different nature to that of a pure KQSL despite being a magnetic state dominated by quantum fluctuations and possibly hosting topologically protected excitations. Our results have implications in the theoretical understanding of the role of bond-disorder within the Kitaev model and in QSL in general. 

\section{Methods}

\subsection{Sample growth and characterization}

Precursor single crystals of \Li{} were grown as described elsewhere \cite{Freund2016}. 40 x 40 um single crystals of \HLIO{} was grown via a topotactic exchange by placing \Li{} in an acidic environment for several hours \cite{bette_solution_2017}. Sample characterization including powder X-ray diffraction, specific heat and magnetization measurements can be found in Ref. \cite{molecules27030871}

\subsection{Resonant Inelastic X-ray Scattering}

RIXS measurements in horizontal scattering geometry were performed at the 27 ID beamline of the Advance Photon Source with $\pi$-incident X-ray polarization. The low-energy RIXS spectra summing over $\pi$-$\sigma^{\prime}$ and $\pi$-$\pi^{\prime}$ polarization channels was collected with a fixed 2 m radius spherically diced Si(844) analyzer positioned to achieve a $[-150,200]$~meV energy window around the elastic line. For the high resolution measurements a second upstream monochromator was used to obtain a resolution of 20~meV. To minimize Thomson scattering the $2\theta$ angle was fixed at $90^{\circ}$. This lead to a small variation of the $L$ values between 5.9 and 5.95 r.l.u. An iris mask was used to achieve a spectrometer momentum resolution of $\pm 0.048~\AA^{-1}$. Each spectra shown is the average of 2-3 scans.

\subsection*{Exact Diagonalization calculations}

We model the crystal field excitations in \HLIO{} and \Li{} by consider a Hamiltonian including spin orbit coupling ($\lambda = 540
$~meV) in the large cubic crystal field limit $H = H_U + H_{CF} + H_{t}$ including on-site Coulomb interactions ($H_U$), trigonal distortions $H_{CF}$ and nearest-neighbors hopping $H_t$. $H_t$ follows the form described in Ref. \cite{delatorre_PRB_2021} and Ref. \cite{delatorre_PRM_2022}. The leading hopping integrals are $t_{||}$, which parameterizes hopping between parallel orbitals and $t_O$ which encodes hopping paths mediated by O $2p$ orbitals. These two hopping integrals are sampled from a normal distribution with standard deviations $\sigma_{t_0} = 50$~meV, $\sigma_{t_{||}} = 10$~meV. For the magnitude of the trigonal fields, $\delta$,  the width of the normal distribution is allowed to vary between \HLIO{} ,$\sigma_{\delta} = 20$~meV, and \Li{}, $\sigma_{\delta} = 5$~meV \cite{delatorre_PRM_2022}. Our calculations explore the phase diagram given by the mean values ($\delta$,$t_O$,$t_{||}$) to find the best agreement with the data. For \HLIO{} we found  
$\delta = -47$~meV, $t_O = 440$~meV and $t_{||} = -50$~meV. For \Li{} this is found for $\delta = -50$~meV, $t_O = 400$~meV and $t_{||} = -30$~meV (See Supplementary Information).

\subsection*{O K edge XAS}

Low temperature O K edge XAS measurements were performed at the beamline 8 of the ALS in both partial fluorescence yield and total electron yield. We fit the data to a combination of two Lorentzians to account for the $\alpha$ peak, two Gaussians for the $\beta$ peak and an arctan step.
 
\section*{Data availability}

Data is available from the authors upon reasonable request.

\section*{Acknowledgements}

We thank J. Knolle and P. A. Lee for useful discussions. Work at Brown University was supported by the U.S. Department of Energy, Office of Science, Office of Basic Energy Sciences, under Award Number DE-SC0021. The work at Boston College was supported by the National Science Foundation under award number DMR-2203512. This research used resources of the Advanced Photon Source, a U.S. Department of Energy (DOE) Office of Science user facility operated for the DOE Office of Science by Argonne National Laboratory under Contract No. DE-AC02-06CH11357. 

\section*{Author Contributions}

A.d.l.T. and B.Z. performed the X-ray spectroscopy measurements with support from M.H.U, J.K., G.F., G.-H. L., W.Y. and D.H. on samples synthesized by F.B and F.T. A.d.l.T. and K.W.P. analyzed and interpreted the data. A.d.l.T. and K.W.P. and wrote the paper with input from all authors.

\section*{Competing Interests}

The authors declare no competing interests.

\renewcommand{\thefigure}{S\arabic{figure}}
\renewcommand{\thetable}{S\arabic{table}}
\renewcommand{\theequation}{S\arabic{equation}}

\pagebreak
\clearpage
\begin{widetext}
\begin{center}
\textbf{\large Supplementary information: Momentum-independent magnetic excitation continuum in the honeycomb iridate H$_3$LiIr$_2$O$_6$}
\end{center}

\section{Incident energy of the RIXS spectra}
\begin{figure*}[!ht]
  \begin{center} \includegraphics[width=\textwidth,clip]{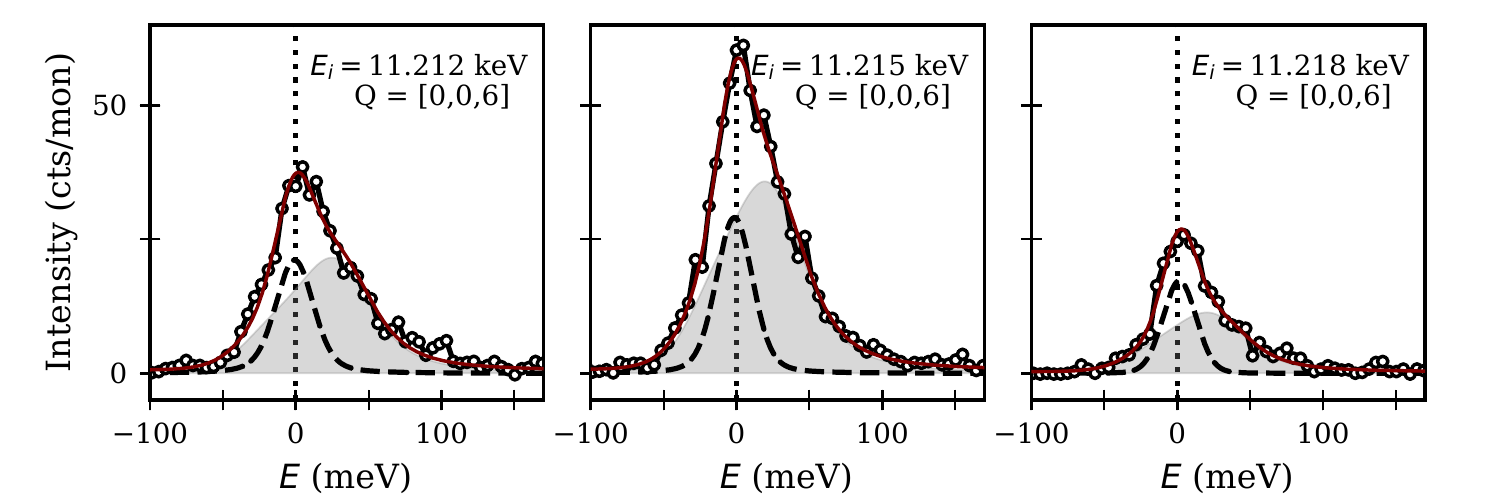}
  \caption{Room temperature resonant inelastic X-ray spectra of \HLIO{} at $Q = [0,0,6]$ at three different values of $E_i$ showing the resonant behavior of the magnetic excitation continuum. Energy resolution for this data set was set to FWHM $ = 28$~meV. Maroon solid line is a fit to a Voigt profile for the elastic intensity (dashed line) and an overdamped harmonic oscillator for the inelastic intensity (grey shading).}
  \label{fig:SI_1}
  \end{center}
\end{figure*}

In Fig. \ref{fig:SI_1}, we show room temperature RIXS spectra of \HLIO{} at $Q = [0,0,6]$ at three different X-ray incident energies. The continuum of magnetic excitations displays resonant behavior, as signified by the change of intensity of the overdamped harmonic oscillator (grey solid area) with incident x-ray energy. The extracted center energy, $E = 38 \pm 5$~meV, and width, $\sigma = 45 \pm 5$~meV are consistent with those found with the higher resolution data shown in Fig. 1 of the main text. 

\section{Details of the crystal field analysis}

\begin{figure*}[!ht]
  \begin{center} \includegraphics[width=\textwidth,clip]{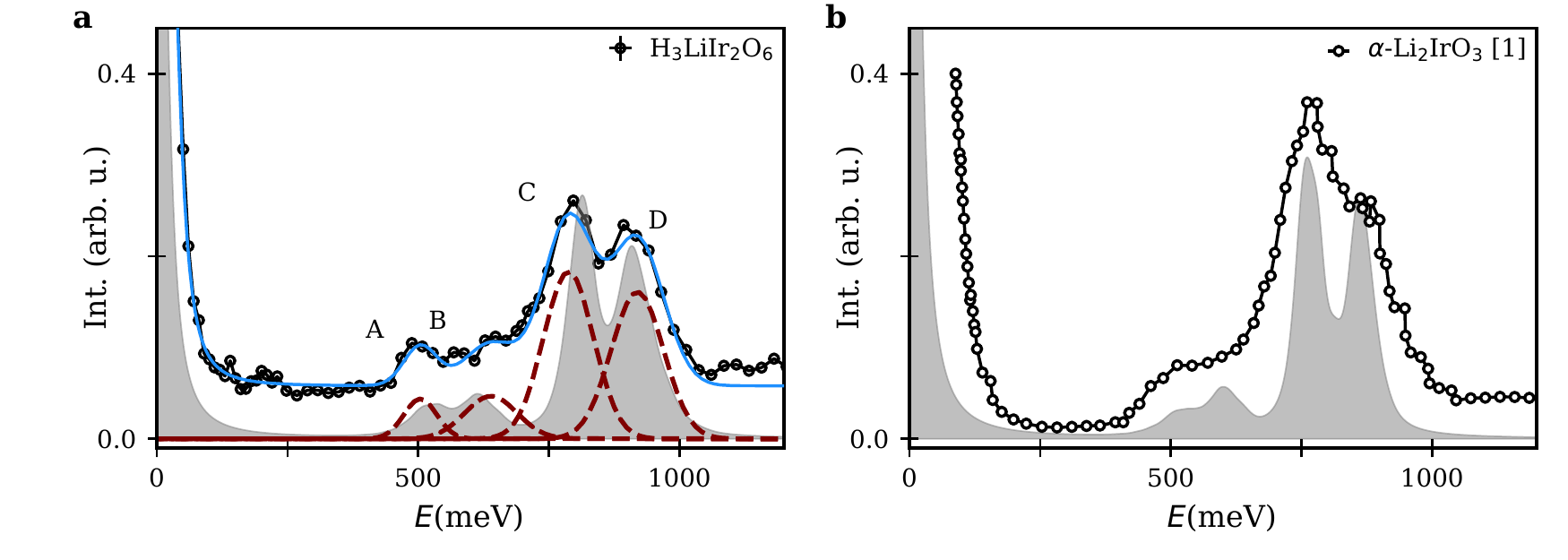}
  \caption{\textbf{a} Intra-$t_{2g}$ RIXS excitations at $E_i = 11.215$~keV (circular markers) compared to the calculated RIXS intensity from exact diagonalization calculations (grey shading). Blue solid line is a fit to the data including a Voigt peak, a DHO, four Gaussian peaks (red dashed lines) and an arctan step to account for the background. \textbf{b} Intra-$t_{2g}$ RIXS excitations at $E_i = 11.215$~keV (circular markers) extracted from \cite{Gretarsson_PRL_2015} compared to the calculated RIXS intensity (grey shading).  }
  \label{fig:SI_2}
  \end{center}
\end{figure*}

In Fig. \ref{fig:SI_2} \textbf{a} we reproduce the data from Fig. 4 \textbf{b} highlighting the four Gaussians peaks(A -D) needed to account for the intra-$t_{2g}$ excitations centered at $E_A = 504.32$~meV, $E_B =639.38$~meV, $E_C = 787.89$ and $E_D = 919.48$~meV. In Fig. \ref{fig:SI_2} \textbf{b} we show RIXS data for \Li{} digitized from Ref. \cite{Gretarsson_PRL_2015}. The grey shaded curve shows the calculated RIXS intensity from the exact diagonalization of a Hamiltonian in the large cubic crystal field limit $H = H_U + H_{CF} + H_{t}$  including spin orbit coupling ($\lambda = 540$~meV), on-site Coulomb interactions ($H_U$), trigonal distortions $H_{CF}$ and nearest-neighbors hopping $H_t$ \cite{delatorre_PRB_2021,delatorre_PRM_2022}. To account for local disorder induced broadening, the trigonal fields $\delta$, and leading hopping integrals $t_O$ and $t_{||}$ are randomly sampled from a normal distribution with standard deviation $\sigma_{t_0} = 50$~meV and $\sigma_{t_{||}} = 10$~meV. The standard deviation for $\delta$ is allowed to vary between \HLIO{}, $\sigma_{\delta} = 20$~meV, and \Li{} $\sigma_{\delta} = 5$~meV. Our calculations explore the phase diagram given by the mean values ($\delta$,$t_O$,$t_{||}$) to find the set of parameters that better account for the observed RIXS intensity. For \HLIO{} we find $\delta = -47$~meV, $t_O = 440$~meV and $t_{||} = -50$~meV and for \Li{} $\delta = -50$~meV, $t_O = 400$~meV and $t_{||} = -30$~meV (See Supplementary Information).

\section{Momentum dependence of crystal field excitations}

\begin{figure}[!ht]
  \begin{center} \includegraphics[width=\textwidth,clip]{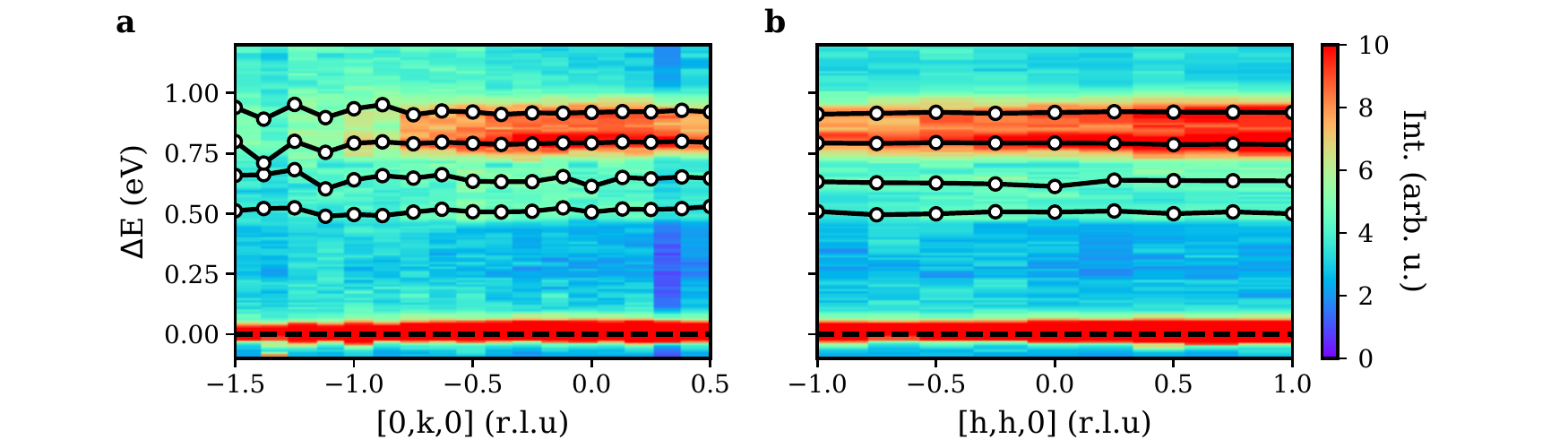}
  \caption{Momentum dependence of the RIXS intensity of the intra-$t_{2g}$ crystal field excitations along \textbf{a} $[0,k,0]$ and \textbf{b} 
  $[h,h,0]$. Circular markers indicate the extract energy from a fit to the data as described in the previous section.}
  \label{fig:SI_4}
  \end{center}
\end{figure}
Fig. \ref{fig:SI_4} \textbf{a} and \textbf{b} shows the intensity variation of the high-energy RIXS spectrum as a function of momentum along $[0,k,0]$ and $[h,h,0]$ respectively. We observed differences of up to a factor of 2 in the inelastic intensity at equivalent momentum points but no dispersion of the crystal field (CF) excitations was observed. The CFs intensity variation is not intrinsic to \HLIO{}. To collect the momentum dependent data shown in Fig. 1, 2 and 3 of the main text and in Fig. \ref{fig:SI_4}, $\theta$, the X-ray incidence angle, $\phi$, the azimuthal angle, and $\chi$, defining the tilt of the sample normal with respect to the scattering plane. Given the sample size ($40\mu$m $\times$ $40\mu$m) comparable to the spot size of the X-ray beam, we cannot exclude that the observed intensity variation is related to a small walk of the sample away from the center of rotation. Another possible extrinsic artifact at play are self-absorption effects dependent on the values of $\theta$ and $2\theta$, the scattering angle, and on the incident and outgoing energy which are known to occur near resonance in RIXS experiments at the Ir $L_3$ edge \cite{revelli_fingerprints_2020}.

\newpage
 \section{X-ray Absorption Spectroscopy measurements in \HLIO{}}
 \begin{figure}[!ht]
  \begin{center} \includegraphics[width=0.6\textwidth,clip]{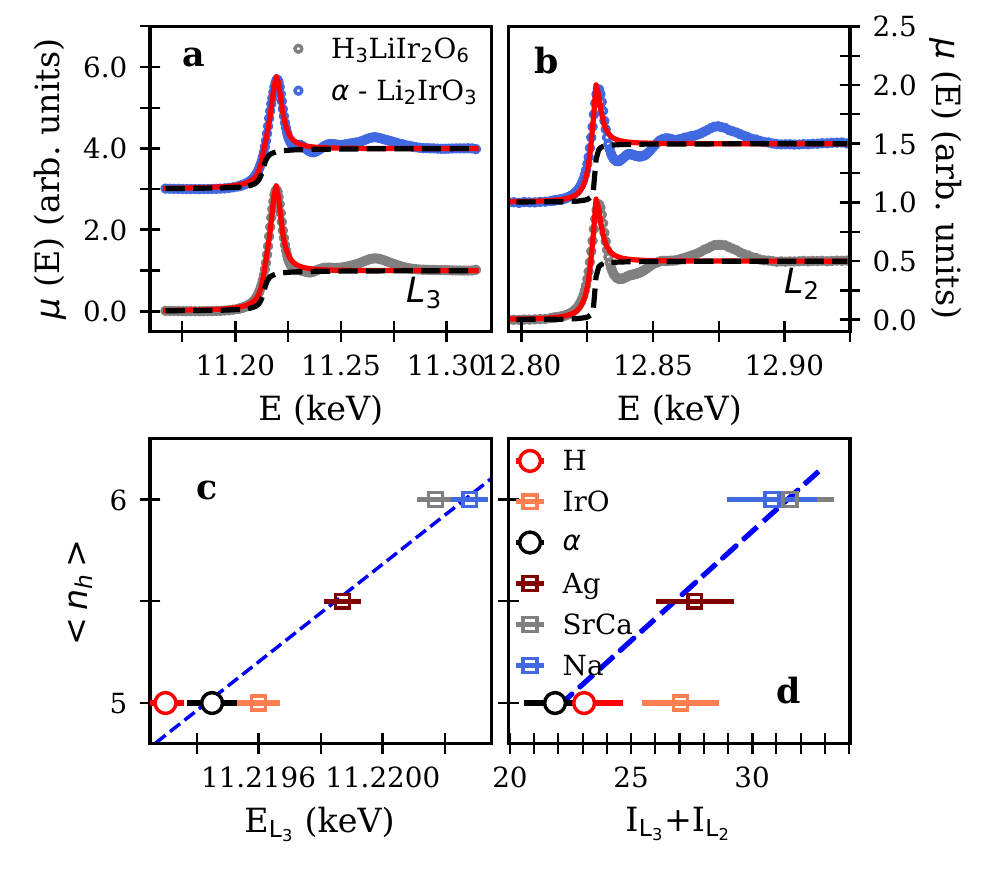}
  \caption{\textbf{a} Ir $L_3$ edge XAS intensity in \HLIO{} (blue markers) and \Li{} (grey markers). \textbf{b} Ir $L_2$ edge XAS intensity in \HLIO{} (blue markers) and \Li{} (grey markers). Red line is a fit to a lo Lorentzian peak and an arctangent step (dotted black line). All data was taken at $T = 300$~K. \textbf{c} Direct comparison of the $L_3$ white line intensity of \HLIO{} (H) to that of \Li{} ($\alpha$) and standard compounds Sr$_3$CaIr$_2$O$_9$ (SrCa), NaIrO$_3$ (Na), Ag$_3$LiIr$_2$O$_6$ (Ag) and IrO. \textbf{d} Same as \textbf{c} but showing the integrated intensity $L_3 + L_2$.}
  \label{fig:SI_5}
  \end{center}
\end{figure}
In Fig. \ref{fig:SI_5} \textbf{a} and\textbf{b}, we show the XAS intensity of \HLIO{} and \Li{} at the Ir $L_3$ and $L_2$ edge, respectively. No change is observed in either the $L_3$ white line position or $L_3 + L_2$ integrated intensity shown in  Fig. \ref{fig:SI_5}\textbf{c} and\textbf{d}. Thus, the introduction of H does not modify the Ir$^{4+}$ oxidation state in \HLIO{}
\end{widetext}

\end{document}